\documentclass[toc]{PoSmodified}

\usepackage{amssymb}
\usepackage{amsmath}
\usepackage{amsfonts}
\usepackage{bbm}
\usepackage{amsthm}
\usepackage{mathrsfs}
\usepackage{color}
\usepackage[all,cmtip]{xy}

\theoremstyle{plain}
\newtheorem{theo}{Theorem}[section]

\theoremstyle{definition}

\newtheorem{ex}[theo]{Example}

\numberwithin{equation}{section}

\def\nn{\nonumber}

\def\hom{\mathrm{hom}}

\def\dd{\mathrm{d}}

\def\id{\mathrm{id}}
\def\ev{\mathrm{ev}}

\def\ra{\triangleright}

\def\bbR{\mathbb{R}}
\def\bbC{\mathbb{C}}

\def\DDD{\mathrm{ad}_{\bullet}}

\newcommand{\obultimes}{\mathbin{\ooalign{$\otimes$\cr\hidewidth\raise0.17ex\hbox{$\scriptstyle\bullet\mkern4.48mu$}}}}
\newcommand{\ostartimes}{\mathbin{\ooalign{$\otimes$\cr\hidewidth\raise0.17ex\hbox{$\scriptstyle\star\mkern4.48mu$}}}}

\newcommand{\obulplus}{\mathbin{\ooalign{$\boxplus$\cr\hidewidth\raise0.295ex\hbox{$\scriptstyle\bullet\mkern4.7mu$}}}}

\newcommand{\bol}[1]{#1}

\def\sk{\vspace{2mm}}

\title{Working with Nonassociative Geometry \\ and Field Theory}

\ShortTitle{Working with Nonassociative Geometry}

\author{\speaker{Gwendolyn E.~Barnes}, Alexander Schenkel and Richard J.~Szabo$^\ast$\vspace{2mm}\\
Department of Mathematics, Heriot-Watt University, Edinburgh, United Kingdom.\vspace{1mm}\\
Maxwell Institute for Mathematical Sciences, Edinburgh, United Kingdom.\vspace{1mm}\\
The Higgs Centre for Theoretical Physics, Edinburgh, United Kingdom.\vspace{2mm}\\
E-mail: \email{geb31@hw.ac.uk}, \email{as880@hw.ac.uk}, \email{R.J.Szabo@hw.ac.uk}}

\abstract{
We review aspects of our formalism for differential geometry on noncommutative and nonassociative spaces which arise from cochain twist deformation quantization of manifolds. We work in the simplest setting of trivial vector bundles and flush out the details of our approach providing explicit expressions for all bimodule operations, and for connections and curvature. As applications, we describe the constructions of physically viable action functionals for Yang-Mills theory and Einstein-Cartan gravity on noncommutative and nonassociative spaces, as first steps towards more elaborate models relevant to non-geometric flux deformations of geometry in closed string theory.
}
\FullConference{Proceedings of the Corfu Summer Institute 2015 "School and Workshops on Elementary Particle Physics and Gravity"\\
		 1--27 September 2015\\
		 Corfu, Greece\\
Preprint: \  {\tt EMPG--16--02}}

\begin{document}


\section{Introduction and summary}

Recent advances in understanding flux compactifications of string
theory have suggested that non-geometric frames are related to
noncommutative and nonassociative deformations of spacetime
geometry~\cite{Blumenhagen:2010hj,Lust:2010iy,Chatzistavrakidis:2012qj,Andriot:2012,Blair:2014kla}; as these flux deformations of geometry are probed by \emph{closed} strings, they have a much better potential for providing an effective target space description of quantum gravity than previous appearances of noncommutative geometry in string theory. In the standard
T-duality orbit $\mathsf{H}\to\mathsf{f}\to\mathsf{Q}\to \mathsf{R}$
relating geometric and non-geometric fluxes, $\mathsf{Q}$-flux
backgrounds experience a noncommutative but strictly associative
deformation while the purely non-geometric $\mathsf{R}$-flux
backgrounds witness a noncommutative and nonassociative
geometry. Nonassociativity in this setting can be encoded by certain
triproducts of fields on configuration
space predicted by off-shell amplitudes in conformal field theory~\cite{Blumenhagen:2011ph} and in double field theory~\cite{Blumenhagen:2013zpa}, or by
nonassociative $\star$-products from deformation
quantization of twisted Poisson structures in the phase space formulation of nonassociative
$\mathsf{R}$-space~\cite{Mylonas:2012pg,Bakas:2013jwa,Mylonas:2013jha}; the
equivalence between these two approaches was demonstrated and extended
in~\cite{Aschieri:2015roa}.
A general treatment of nonassociative
$\star$-products in this context can be found
in~\cite{Kupriyanov:2015dda} (see also the contribution of
V.~Kupriyanov to these proceedings). Reviews of noncommutativity and
nonassociativity in non-geometric closed string theory can be found
in~\cite{Lust:2012fp,Plauschinn:2012kd,Mylonasa,Blumenhagen:2014sba} (see also the
contributions of P.~Schupp and I.~Bakas to these proceedings).
\sk

The cochain twist deformation quantization techniques originally developed
by~\cite{Mylonas:2013jha} were motivated by the search for a
systematic way to generalize notions of differential geometry to such
non-geometric backgrounds, and in particular to construct
nonassociative deformations of field theory and ultimately gravity
(see also~\cite{Aschieri:2015roa}); this approach is different in
spirit to the nonassociative twist deformation of the geometric
$\mathsf{f}$-flux frame considered in~\cite{D'AF}, which does not seem
to be of relevance for non-geometric string theory, nor does it agree
with the string theory inspired nonassociative torus bundles
of~\cite{Bouwknegt:2004ap,Hannabuss:2010tp} which reproduce the
classical limit only up to Morita equivalence. Physically consistent
models with novel properties in the context of quantum
mechanics were constructed in~\cite{Mylonas:2013jha} using this
formalism, and of Euclidean scalar quantum field theory
in~\cite{Mylonas:2014kua}. To extend these considerations to more
complicated field theories, a general systematic formalism was
developed in~\cite{Barnes:2014,Barnes:2015} for differential geometry on
noncommutative and nonassociative spaces internal to the
representation category of any quasi-Hopf algebra, generalizing and extending earlier work~\cite{Bouwknegt:2007sk,BeggsMajid1,Aschieri:2012ii}. This is the starting point for the present contribution. 
\sk

The purpose of this contribution is to unpack and make explicit the somewhat
abstract categorical constructions of~\cite{Barnes:2014,Barnes:2015} in a less formal language that we hope will be palatable to a larger audience. We focus
on the special case of most physical relevance: the cochain twist quantization
of a classical manifold; this construction is reviewed in
Section~\ref{sec:NASV}. The formalism is powerful enough to capture
the cases of constant non-geometric fluxes as well as non-constant
ones such as those which arise in the flux formulation of double
field theory~\cite{Blumenhagen:2013zpa}; in fact, our constructions in the remainder of
this paper
are completely general and can be applied to a much broader framework
without specific reference to string theory. We further restrict to
trivial vector bundles over these noncommutative and nonassociative spaces with
diagonal action of the pertinent Hopf algebra of symmetries of the
non-geometric background. This simplification enables us to give very explicit ``local''
descriptions of the noncommutative and nonassociative geometry while still retaining
generic features and indicating how the general formalism
of~\cite{Barnes:2014,Barnes:2015} may be applied to constructions of
physically viable field theories; in particular, we give concrete realizations of the pertinent bimodule operations for homomorphism bundles. In Section~\ref{sec:NACC} we apply
this framework to obtain explicit expressions for connections
and their curvatures on noncommutative and nonassociative vector bundles. As a starting point for
building more elaborate models describing the low-energy effective dynamics
of closed strings in non-geometric backgrounds, in
Section~\ref{sec:NAFT} we demonstrate how to apply our formalism to
the constructions of physically
sensible action functionals for Yang-Mills theory and Einstein-Cartan gravity on
noncommutative and nonassociative spaces.


\section{Nonassociative spaces and vector bundles\label{sec:NASV}}

\subsection{Spaces}
We briefly review how a classical manifold may 
be deformed into a noncommutative and nonassociative
space by using cochain twist deformation techniques. 
Recall that associated to any manifold $M$ is the Lie algebra
$\mathrm{Vec}(M)$ of vector fields on $M$ (with Lie bracket $[\,\cdot\,,\,\cdot\,]$ 
given by the vector field commutator), 
which plays the role of the infinitesimal diffeomorphisms of $M$. 
This Lie algebra gives rise to a Hopf algebra $U\mathrm{Vec}(M)$, the universal enveloping algebra
of $\mathrm{Vec}(M)$, which is characterized as follows: As an algebra, $U\mathrm{Vec}(M)$ is the free
unital algebra generated by $\mathrm{Vec}(M)$ modulo the relations
$v\,w - w\,v = [v,w]$, for all $v,w\in \mathrm{Vec}(M)$.
The coproduct $\Delta$, counit $\epsilon$ and antipode $S$
on $U\mathrm{Vec}(M)$ are defined on generators by
\begin{subequations}\label{eqn:Hacts}
\begin{flalign}
\Delta(v) &= v \otimes 1 + 1 \otimes v~,\quad \Delta(1) = 1 \otimes 1~, \\[4pt]
\epsilon(v) &= 0~,\quad \epsilon(1) = 1~,\\[4pt]
S(v) &= -v~,\quad S(1) = 1~,
\end{flalign}
\end{subequations}
for all $v \in \mathrm{Vec}(M)$. The maps $\Delta$ and $\epsilon$ are extended as algebra homomorphisms
and $S$ as an anti-algebra homomorphism to all of $U\mathrm{Vec}(M)$.
\sk

In the following we fix a choice of sub-Hopf algebra $H \subseteq U\mathrm{Vec}(M)$, 
which we shall interpret as the symmetries of $M$ along which we want to perform the
deformation quantization. See Examples~\ref{ex:abelian} and~\ref{ex:nonassociative} below for typical choices.
\sk

Let us denote by $A := C^\infty(M)$ the algebra of complex-valued smooth functions on $M$.
The action of vector fields on $A$ as derivations can be extended to
an $H$-action $\ra : H\otimes A\to A$, which preserves the product and
unit in $A$, i.e.\
\begin{flalign}\label{eqn:Leib}
h \ra (a \, b) = \big(h_{(1)} \ra a\big) \, \big(h_{(2)} \ra b\big)~,\quad h\ra 1 =\epsilon(h)\,1~,
\end{flalign}
for all $h\in H$ and $a,b \in A$. Here we have used the 
Sweedler notation $\Delta(h) = h_{(1)} \otimes h_{(2)}$ 
(with summations understood) to abbreviate the coproduct.
In technical terms \eqref{eqn:Leib} states that $A$ is an $H$-module algebra.
\sk

The commutative and associative algebra $A$ can be deformed by using a cochain twist
$F$ of $H$ into a noncommutative and nonassociative algebra $A_\star$.
Recall that a cochain twist is an invertible element $F  = F^{(1)}\otimes F^{(2)} \in H \otimes H$ 
(with summations understood) satisfying the normalization condition
\begin{subequations} \label{eqn:norm}
\begin{flalign}
\epsilon(F^{(1)})\,F^{(2)} &= 1 =  F^{(1)} \, \epsilon(F^{(2)})~.
\end{flalign}
As a consequence, the inverse twist $F^{-1} = F^{(-1)}\otimes F^{(-2)}\in H\otimes H$ (with summations understood)
satisfies a similar normalization condition
\begin{flalign}
\epsilon(F^{(-1)})\,F^{(-2)} &= 1 =  F^{(-1)} \, \epsilon(F^{(-2)})~.
\end{flalign}
\end{subequations}

Given any cochain twist $F \in H \otimes H$, we can deform
the Hopf algebra $H$ into a {\em quasi}-Hopf algebra $H_F$:
As algebras, $H_F$ is the same as $H$ and also the counit
of $H_F$ agrees with that of $H$, i.e.\ $\epsilon_F=\epsilon$. However, the coproduct,
quasi-antipode and associator in $H_F$ are deformed according to
\begin{subequations}
\begin{flalign}
\Delta_F(\,\cdot\,) &:= F\,\Delta(\,\cdot\,)\,F^{-1}~,\\[4pt]
 S_F^{} &:= S~~,\quad \alpha_F :=S(F^{(-1)})\,\alpha\,F^{(-2)}~~,\quad \beta_F:= F^{(1)}\,\beta\,S(F^{(2)})~,\\[4pt]
 \label{eqn:defasso}\phi_F^{} &:= (1\otimes F)\, (\id^{}_H\otimes \Delta)(F)\, \phi \, (\Delta\otimes \id^{}_H)(F^{-1})\, (F^{-1}\otimes 1)~,
\end{flalign}
\end{subequations}
where $\alpha = 1 = \beta$ and $\phi=1\otimes 1\otimes 1$ in the original Hopf algebra $H$.
\sk

The cochain twist $F$ can be used to deform the product $\mu$ 
in the algebra $A$ to a noncommutative and nonassociative $\star$-product
\begin{flalign}\label{eqn:star}
\mu_\star := \mu \circ F^{-1}~.
\end{flalign} 
We denote the resulting noncommutative and nonassociative algebra by $A_\star$
and abbreviate the $\star$-product as $a\star b:= \mu_\star(a\otimes b)$, for $a,b\in A_\star$.
In the spirit of noncommutative geometry, 
we interpret the algebra $A_\star$ as (the algebra of functions on) a noncommutative and nonassociative space.
\sk

By construction, the original $H$-action $\ra : H\otimes A\to A$  
induces an $H_F$-action $\ra : H_F\otimes A_\star\to A_\star$, which preserves the product 
and unit in $A_\star$, i.e.\
\begin{flalign}\label{eqn:Leibequiv}
h \ra (a \star b) = \big(h_{(1)_F} \ra a\big) \star  \big(h_{(2)_F} \ra b\big)~,\quad h\ra 1 =\epsilon_F(h)\,1~,
\end{flalign}
for all $h \in H_F$ and $ a, b \in A_\star $. 
Here we have used the Sweedler notation $\Delta_F^{}(h) = h_{(1)_F} \otimes h_{(2)_F}$ 
(with summations understood) to abbreviate the deformed coproduct.
In technical terms \eqref{eqn:Leibequiv} states that $A_\star$ is an $H_F$-module algebra.
\sk

It is important to observe that the noncommutativity of $A_\star$ 
is controlled by the triangular $R$-matrix
\begin{flalign}
R_F = F_{21}\,R\, F^{-1} = R_F^{(1)} \otimes R_F^{(2)}
\end{flalign}
in $H_F \otimes H_F$, where $R=1\otimes1$ in the original Hopf algebra
$H$ and $F_{21} = F^{(2)} \otimes F^{(1)}$ is the twist 
with flipped legs. Explicitly, the $\star$-product is commutative up to  
the action of $R_F$, i.e.\
\begin{flalign}\label{eqn:flip} 
a \star b = \big(R_F^{(2)} \ra b\big) \star \big(R_F^{(1)} \ra a\big)~,
\end{flalign}
for all $a, b \in A_\star$.
Similarly, the nonassociativity of $A_\star$ is controlled by 
the associator $\phi_F=  \phi_F^{(1)}\otimes  \phi_F^{(2)}\otimes  \phi_F^{(3)}$
in $H_F \otimes H_F\otimes H_F$ given by \eqref{eqn:defasso}.
Explicitly,  the $\star$-product is associative up to  
the action of $\phi_F$, i.e.\
\begin{flalign}
\label{eqn:naA}(a \star b) \star c = (\phi_F^{(1)}\ra a)\star \big((\phi_F^{(2)}\ra b)\star (\phi_F^{(3)}\ra c)\big)~,
\end{flalign}
for all $a,b,c \in A_\star$.

\begin{ex}\label{ex:abelian}
Let $M = \bbR^{m}$ and consider the Abelian cocycle twist (with summation over $ i,j,\dots $ understood here and in the following)
\begin{flalign}
F = \exp\big(- \mbox{$\frac{\mathrm{i}\, \hbar}2$} \, \Theta^{ij}\, P_i\otimes P_j\big)
\end{flalign}
based on the cocommutative Hopf algebra $H= U\mathfrak{g}$,
where $\mathfrak{g}$ is the Abelian Lie algebra of infinitesimal translations $\{P_i: 1\leq i\leq m\}$ 
and $\Theta = (\Theta^{ij})_{i,j=1}^{m}= (\mathsf{Q}^{ij}{}_k\, w^k)_{i,j=1}^{m}$ is an antisymmetric real-valued 
$m\times m$-matrix which arises from a constant
non-geometric $\mathsf{Q}$-flux of closed string theory~\cite{Condeescu:2012sp,Bakas:2015gia}. In this example we have 
\begin{flalign}
R_F =F^{-2} = \exp\big( \mbox{$\mathrm{i}\, \hbar$} \, \Theta^{ij}\, P_i\otimes P_j\big)\quad ,\qquad \phi_F = 1\otimes 1\otimes 1~.
\end{flalign}
In particular $A_\star$ is strictly associative for this choice of twist.
\end{ex}

\begin{ex}\label{ex:nonassociative}
Let $M = \bbR^{2n}=\bbR^{n}\times \bbR^{n} $ and consider the non-Abelian cochain twist 
\begin{flalign}
F=\exp\Big(\mbox{$-\frac{\mathrm{i}\, \hbar}2$}\, \big(\mbox{$\frac14$} \, \mathsf{R}^{ijk}\,
(M_{ij}\otimes P_k-P_i\otimes M_{jk})+P_i\otimes\tilde P\,^i-\tilde
P\,^i\otimes P_i\big)\Big)
\end{flalign} 
based on the cocommutative Hopf algebra $H= U\mathfrak{g}$, 
where $\mathfrak{g}$ is the non-Abelian nilpotent Lie algebra of infinitesimal translations 
and Bopp shifts $\{P_i,\tilde P\,^i,M_{ij} : 1\leq i < j \leq n\}$; the nontrivial 
Lie bracket relations are given by $[\, \tilde P\,^i,M_{jk}] = \delta^{i}{}_{j}\, P_k-\delta^{i}{}_{k}\, P_j $.
Here $\mathsf{R}=(\mathsf{R}^{ijk})_{i,j,k=1}^n$ is a completely antisymmetric 
real-valued tensor of rank~$3$ which arises from a constant
non-geometric $\mathsf{R}$-flux of closed string theory~\cite{Mylonas:2013jha}.
In this example we have 
\begin{flalign}
R_F =F^{-2} \quad ,\qquad \phi_F = \exp\big(\mbox{$\frac{\hbar^2}2$}\, \mathsf{R}^{ijk}\, P_i\otimes P_j\otimes P_k\big)~.
\end{flalign}
In particular $A_\star$ is {\em not} strictly associative for this choice of twist.
\end{ex}


\subsection{Vector bundles}
Given any (complex) vector bundle $E\to M$ over the manifold $M$,
we can consider its smooth sections ${\mit\Gamma}^\infty(E)$, 
which is a bimodule over $A=C^\infty(M)$ with respect to the usual pointwise
module structures. To simplify our considerations in this paper, 
we assume that $E\to M$ is a trivial complex vector bundle of rank $n$, i.e.\ $E = M \times \bbC^n\to M$
with bundle projection given by projecting on the first factor. For a discussion of 
generic vector bundles see \cite{Barnes:2014,Barnes:2015}. 
\sk

The sections of a trivial vector bundle over $M$ of rank $n$
can be described by a free $A$-bimodule $V = A^n$.
Elements $v\in V$ are thus given by column vectors with entries in $A$, i.e.\
\begin{flalign}
v = \begin{pmatrix}
v^1\\\vdots\\v^n
\end{pmatrix}\quad , \qquad v^i\in \bol{A}~,~~ i=1,\dots,n~.
\end{flalign}
Alternatively, we can make use of the standard basis
$\{e_i\}_{i=1}^n$ and write
\begin{flalign}
v = e_i \, v^i\quad , \qquad v^i\in \bol{A}~,~~ i=1,\dots,n~.
\end{flalign}
The left and right $\bol{A}$-actions on $V$ are given componentwise, i.e.\
\begin{subequations}
\begin{flalign}
a \,\, v &:= e_i \, (a\,\,v^i) ~,\\[4pt]
v\,\,a &:= e_i \, (v^i\,\,a)~,
\end{flalign}
\end{subequations}
for all $a\in \bol{A}$ and $v \in V$. 
Similarly, we equip $V$ with a componentwise
$H$-action $\ra : H\otimes V\to V$, i.e.\
\begin{flalign}
h\ra v &:= e_i \, (h \ra v^i)~,
\end{flalign}
for all $h \in H$ and $v \in V$. 
It follows that 
\begin{flalign}
h \ra e_i  = \epsilon(h) \, e_i~,
\end{flalign}
for all $h \in H$ and $i = 1, \dots, n$, 
i.e.\ the basis $\{e_i\}_{i=1}^n$ is $H$-invariant. 
As a consequence of  \eqref{eqn:Leib}, we obtain further that
\begin{subequations}\label{eqn:hVcomp}
\begin{flalign}
h \ra (a \,\, v) &= \big(h_{(1)} \ra a\big)\, \big(h_{(2)} \ra v\big) ~,\\[4pt]
h \ra(v\,\,a) &= \big(h_{(1)} \ra v\big)\,\,\big(h_{(2)} \ra a\big)~,
\end{flalign}
\end{subequations}
for all $ a \in A$, $v \in V $ and $h\in H$. 
In technical terms \eqref{eqn:hVcomp} states that
$V$ is an $H$-module bimodule over the $H$-module algebra $A$.
\sk

We have explained how a twist
$F\in H\otimes H$ can be used 
to deform the Hopf algebra $H$ to a quasi-Hopf algebra $H_F$,
and the commutative and associative algebra $A$ to a noncommutative 
and nonassociative algebra $A_\star$. Similarly, we can deform $V$ into
an $H_F$-module $A_\star$-bimodule $V_\star$ by introducing
the $H_F$ and $A_\star$-actions
\begin{subequations}
\begin{flalign}
\label{eqn:Hcomponent} h\ra v &:= e_i \, (h \ra v^i)~,\\[4pt]
\label{eqn:leftA} a \,\star\, v &:= e_i \, (a\,\star\,v^i)~,\\[4pt]
\label{eqn:rightA} v\,\star\,a &:= e_i \, (v^i\,\star\,a)~,
\end{flalign}
\end{subequations}
for all $h\in H_F$, $a\in \bol{A}_\star$ and $v \in V_\star$. 
One easily verifies the compatibility conditions between the $H_F$ and $A_\star$-actions
\begin{subequations}
\begin{flalign}
h \ra (a \,\star\, v) &= \big(h_{(1)_F} \ra a\big)\,\star\, \big(h_{(2)_F} \ra v\big) ~,\\[4pt]
h \ra(v\,\star\,a) &= \big(h_{(1)_F} \ra v\big)\,\star\,\big(h_{(2)_F} \ra a\big)~,
\end{flalign}
\end{subequations}
for all $ h \in H_F $, $a\in A_\star$ and $v\in V_\star$. In the spirit of noncommutative geometry, we interpret $V_\star$ as (the module of sections of) 
a vector bundle over $ A_\star $.
\sk

Noncommutativity of the $A_\star$-bimodule structure 
is controlled as in \eqref{eqn:flip} by the $R$-matrix $R_F$, i.e.\ 
\begin{subequations}\label{eqn:sym}
\begin{flalign}
a \star v &= \big(R_F^{(2)} \ra v\big)\,\star\,\big(R_F^{(1)} \ra a\big)~, \\[4pt]
v \star a &= \big(R_F^{(2)} \ra a\big)\,\star\,\big(R_F^{(1)} \ra v\big)~,
\end{flalign}
\end{subequations}
for all $ a \in A_\star$ and $ v \in V_\star $, while
nonassociativity is controlled as in \eqref{eqn:naA} by the associator $\phi_F$, i.e.\
\begin{subequations}\label{eqn:nalractions}
\begin{flalign}
(a \star b) \star v &= (\phi_F^{(1)}\ra a)\star \big((\phi_F^{(2)}\ra b)\star (\phi_F^{(3)}\ra v)\big)~,\\[4pt]
v \star (a \star b) &= \big((\phi_F^{(-1)}\ra v)\star (\phi_F^{(-2)}\ra a)\big)\star (\phi_F^{(-3)}\ra b)~,
\end{flalign}
\end{subequations}
for all $ a,b \in A_\star$ and $ v \in V_\star $.
Here we have denoted the components of the inverse associator by 
$\phi_F^{-1} = \phi_F^{(-1)}\otimes \phi_F^{(-2)}\otimes \phi_F^{(-3)}$
(with summations understood).


\subsection{Homomorphism bundles}
Many interesting objects in differential geometry 
are described by maps between vector bundles.
For example, a metric is a map $g : TM \to T^\ast M$ from the tangent bundle to the cotangent bundle, while the curvature of a connection on a vector bundle $E\to M$
is a map $ E \to E\otimes \bigwedge^2 T^\ast M$.
Recall that vector bundle maps between two vector bundles $E \to M$ and $E^\prime\to M$
can be equivalently described by sections of the homomorphism bundle
$\hom(E,E^\prime)\to M$. The module of sections
${\mit\Gamma}^\infty(\hom(E,E^\prime))$ of the homomorphism bundle
is isomorphic (as a $C^\infty(M)$-bimodule) to the module of right module maps
$\hom_{C^\infty(M)} ({\mit\Gamma}^\infty(E),{\mit\Gamma}^\infty(E^\prime))$; the latter 
are linear maps $L : {\mit\Gamma}^\infty(E)\to {\mit\Gamma}^\infty(E^\prime)$ 
which satisfy additionally the right $C^\infty(M)$-linearity condition
\begin{flalign}\label{eqn:rightAlinclassical}
L(v\,a) = L(v)\,a~,
\end{flalign}
 for all $v\in {\mit\Gamma}^\infty(E)$ and $a\in C^\infty(M)$.
\sk

Our goal now is to describe the analog of homomorphism
bundles in our noncommutative and nonassociative framework.
Given  two modules $V_\star =A_\star^n$ and $W_\star =A_\star^m$,
we first consider the vector space of linear maps $\hom_F^{}(V_\star,W_\star)$
from $V_\star$ to $W_\star$. This vector space comes together with a natural
$H_F$-action $\ra : H_F\otimes \hom_F^{}(V_\star,W_\star) \to \hom_F^{}(V_\star,W_\star)$
given by the adjoint action
\begin{flalign}\label{eqn:adjointHF}
h\ra L := \big(h_{(1)_F}\ra \,\cdot\,\big)\circ L \circ \big(S_F(h_{(2)_F}) \ra\,\cdot\, \big)~,
\end{flalign}
for all $h\in H_F$ and $L\in \hom_F^{}(V_\star,W_\star)$. It is important to stress 
that we {\em do not} require the linear maps $L : V_\star \to W_\star$ to preserve the $H_F$-action.
As explained in~\cite[Section~1]{Barnes:2014}, this would lead to an overly rigid
framework for studying noncommutative and nonassociative geometry.
\sk

The standard operations of evaluating linear maps $\hom_F^{}(V_\star,W_\star)$
on elements in $V_\star$ and composing or tensoring linear maps with each other 
are in general not compatible with the $H_F$-action given in \eqref{eqn:adjointHF}.
In particular, for generic cochain twists $F$ we have the {\em non-equality}
\begin{flalign}\label{eqn:inequality}
h\ra \big(L(v)\big) \neq \big(h_{(1)_F}\ra L\big)\big(h_{(2)_F}\ra v\big)~,
\end{flalign}
for some $h\in H$, $L\in \hom_F^{}(V_\star,W_\star)$ and $v\in V_\star$.
Using internal homomorphism techniques from category theory, 
one can show that there exist deformations of the evaluation,
composition and tensor product operations
which are compatible with the $H_F$-actions~\cite{Barnes:2014}. We
denote these by
\begin{subequations}\label{eqn:operationsgeneral}
\begin{flalign}
\ev_F^{} &: \hom_F^{}(V_\star,W_\star) \otimes_\star V_\star \longrightarrow W_\star~, \label{eqn:evaluationgeneral}\\[4pt]
\bullet_F^{} &: \hom_{F}^{} (W_\star,X_\star) \otimes_\star \hom_{F}^{}(V_\star,W_\star) \longrightarrow \hom_{F}^{}(V_\star,X_\star)~, \label{eqn:compgeneral}\\[4pt]
\obultimes_F^{} &: \hom_F^{} (V_\star,X_\star) \otimes_\star \hom_F^{}(W_\star,Y_\star) \longrightarrow 
\hom_F^{} (V_\star \otimes_\star W_\star,X_\star \otimes_\star Y_\star)~, \label{eqn:tensorgeneral}
\end{flalign}
\end{subequations}
and refer to~\cite{Barnes:2014,Barnes:2015} for further details. The $\star$-tensor product 
$V_\star\otimes_{\star} W_\star$ is the ordinary tensor product of vector spaces
equipped with the $H_F$-action 
\begin{flalign}
h \ra (v\otimes_{\star} w) = \big(h_{(1)_F} \ra v\big)\otimes_{\star} \big(h_{(2)_F}\ra w\big)~,
\end{flalign}
for all $h\in H_F$, $v\in V_\star$ and $w\in W_\star$.
For the example of the evaluation $\ev_F^{}$, compatibility with the $H_F$-actions means that
\begin{flalign}
h\ra \ev_F^{} (L\otimes_\star v) = \ev_F^{}\big((h_{(1)_F}\ra L) \otimes_\star (h_{(2)_F}\ra v)\big)~,
\end{flalign}
for all $h\in H$, $L\in \hom_F^{}(V_\star,W_\star)$ and $v\in V_\star$, 
which resolves the problem encountered in \eqref{eqn:inequality}.
\sk

The $H_F$-compatible version of the right $A$-linearity condition \eqref{eqn:rightAlinclassical} 
is given by the weak right $A_\star$-linearity condition
\begin{flalign}\label{eqn:weakright}
\ev_F^{} \big(L\otimes_{\star} (v\,\star\,a)\big) =  \ev_F^{}\big((\phi_F^{(-1)}\ra L) \otimes_{\star} (\phi_F^{(-2)}\ra v) \big)\,\star\,(\phi_F^{(-3)}\ra a)~,
\end{flalign}
for all $v\in V_\star$ and $a\in A_\star$. We denote by $\hom_{A_\star}(V_\star,W_\star)$
the vector space of all linear maps $L\in \hom_F^{}(V_\star,W_\star)$ which satisfy the condition 
\eqref{eqn:weakright}. It can be shown that $\hom_{A_\star}(V_\star,W_\star)$ is 
an $H_F$-module $A_\star$-bimodule, and hence a noncommutative and nonassociative vector bundle
in its own right. We interpret $\hom_{A_\star}(V_\star,W_\star)$ as (the module of sections of)
the homomorphism bundle from $V_\star$ to $W_\star$. Moreover, the operations \eqref{eqn:operationsgeneral}
induce to 
\begin{subequations}\label{eqn:operationsgeneralA}
\begin{flalign}
\ev_{A_\star}^{} &: \hom_{A_\star}^{}(V_\star,W_\star) \otimes_{A_\star} V_\star \longrightarrow W_\star~, \label{eqn:evaluationgeneralA}\\[4pt]
\bullet_{A_\star}^{} &: \hom_{A_\star}^{} (W_\star,X_\star) \otimes_{A_\star} \hom_{A_\star}^{}(V_\star,W_\star) 
\longrightarrow \hom_{A_\star}^{}(V_\star,X_\star)~, \label{eqn:compgeneralA}\\[4pt]
\obultimes_{A_\star}^{} &: \hom_{A_\star}^{} (V_\star,X_\star) \otimes_{A_\star} \hom_{A_\star}^{}(W_\star,Y_\star)
 \longrightarrow \hom_{A_\star}^{} (V_\star \otimes_{A_\star} W_\star,X_\star \otimes_{A_\star} Y_\star)~, \label{eqn:tensorgeneralA}
\end{flalign}
\end{subequations}
where $\otimes_{A_\star}$ denotes the tensor product over $A_\star$. Explicitly, 
$V_\star\otimes_{A_\star} W_\star$ is the quotient of $V_\star \otimes_\star W_\star$ 
by the relations
\begin{flalign}\label{eqn:quotient}
(v\,\star\,a)\otimes_\star w = (\phi_F^{(1)}\ra v)\otimes_\star \big((\phi_F^{(2)}\ra a)\,\star\,(\phi_F^{(3)}\ra w)\big)~,
\end{flalign}
for all $a\in\bol{A}_\star$, $v\in\bol{V}_\star$ and $w\in\bol{W}_\star$.
\sk

As $V_\star = A_\star^n$ and $W_\star = A_\star^m$ 
are by assumption free $A_\star$-bimodules (as are $X_\star$ and $Y_\star$), we can make use of the corresponding bases 
$\{e_i\}_{i=1}^n$ and $\{e_j\}_{j=1}^m$
to find simple expressions 
for the homomorphisms  $\hom_{A_\star}(V_\star,W_\star)$, and in particular
the operations \eqref{eqn:operationsgeneralA}. In the following, we shall denote
(with an abuse of notation) all bases by the same symbols.

\paragraph{Evaluation:}
Because of the weak right $A_\star$-linearity condition 
\eqref{eqn:weakright}, any $L\in \hom_{A_\star}(V_\star,W_\star)$
is specified by its evaluation on the basis $\{e_i\}_{i=1}^n$ of $V_\star$.
Using also the basis $\{e_j\}_{j=1}^m$ of $W_\star$, we have the expansion
\begin{flalign}\label{eqn:evchar}
\ev_{A_\star}^{}(L \otimes_{A_\star} e_i) = e_j \, {L^j}_i~,
\end{flalign}
which allows us to characterize $L$ in terms of an $m\times n$-matrix with coefficients
given by ${L^j}_i \in A_\star$. Hence we have established an isomorphism of vector spaces
\begin{flalign}\label{eqn:homAisomatrix}
\hom_{A_\star}(V_\star,W_\star) \longrightarrow  A_\star^{m\times n}~,\quad L\longmapsto ({L^{j}}_{i})~,
\end{flalign}
which assigns to any $L$ its matrix representation.
The evaluation of $L \in \hom_{A_\star}(V_\star,W_\star)$
on a generic element $v = e_i \, v^i \in V_\star$ can then be expressed as
\begin{flalign}\label{eqn:ev}
\nn \ev_{A_\star}^{}(L \otimes_{A_\star} v) &= \ev_{A_\star}^{}(L
\otimes_{A_\star} (e_i \, v^i)) \\[4pt]
\nn &= \ev_{A_\star}^{}\big((\phi_F^{(-1)}\ra L) \otimes_{A_\star} (\phi_F^{(-2)}\ra e_i) \big)\,\star\,(\phi_F^{(-3)}\ra v^i)\\[4pt] 
\nn &= \ev_{A_\star}^{}\big(L \otimes_{A_\star}  e_i \big)\,\star\, v^i
\\[4pt] &= (e_j \, {L^j}_i) \,\star\, v^i = e_j \, ({L^j}_i \,\star\, v^i) ~.
\end{flalign}
In the second step  we have used \eqref{eqn:weakright}
and $e_i \, v^i = e_i\star v^i$, which follows from $H_F$-invariance of
the basis and normalization of the twist.
The third step follows by using again $H_F$-invariance of
the basis and also normalization of the associator.
\sk

Because the evaluation operation is compatible with the $H_F$-actions,
it follows that
\begin{flalign}\label{eqn:HonLbasis}
\ev_{A_\star}^{} \big((h \ra L) \otimes_{A_\star} e_i\big) = 
\ev_{A_\star}^{} \big(h\ra (L\otimes_{A_\star} e_i)\big)=
h \ra \ev_{A_\star}^{}\big( L \otimes_{A_\star} e_i\big)= 
 e_j \, \big(h \ra {L^j}_i\big) ~,
\end{flalign}
for all $h \in H_F$ and  $ L \in\hom_{A_\star}^{}(V_\star,W_\star)$,
where in the first step we have used again $H_F$-invariance of the basis. 
It follows that,
by equipping $A_\star^{m\times n}$ with the componentwise $H_F$-action,
the isomorphism \eqref{eqn:homAisomatrix} is an 
isomorphism of $H_F$-modules. By equipping $A_\star^{m\times n}$ further with the componentwise $A_\star$-bimodule
structure, the map \eqref{eqn:homAisomatrix} is an isomorphism of $H_F$-module $A_\star$-bimodules.

\paragraph{Composition:} 
Given $ V_\star = A_\star^n$, $W_\star = A_\star^m$ and $X_\star = A_\star^l$,  
one can show by similar calculations 
that the composition $L^\prime\bullet_{A_{\star}} L\in \hom_{A_\star}(V_\star,X_\star)$
of any $L\in \hom_{A_\star}(V_\star,W_\star)$ and $L^\prime \in \hom_{A_\star}(W_\star,X_\star)$
is given by the components
\begin{flalign}\label{eqn:comp}
\ev_{A_\star}^{} \big(\big(L^\prime \bullet_{A_\star}^{} L \big)\otimes_{A_\star} e_i\big) 
= e_k \, \big( {{L^\prime}\,^k}_j \,\star\, {L^j}_i\big)~.
\end{flalign}
Hence the isomorphism \eqref{eqn:homAisomatrix} sends the composition operation 
$\bullet_{A_{\star}}$ to the $\star$-matrix product 
\begin{flalign}\label{eqn:compmatrix}
\star : A_\star^{l\times m}\otimes_{A_\star} A_\star^{m\times n}
\longrightarrow A_\star^{l\times n}~,\quad ({{L^\prime}\,^k}_j) \otimes_{A_\star} ({{L}^j}_i)\longmapsto
({{L^\prime}\,^k}_j \,\star\, {L^j}_i)~.
\end{flalign}
In the special case where $V_\star = W_\star =X_\star$, it follows that the
endomorphism algebra  $\mathrm{end}_{A_\star}(V_\star):= \hom_{A_\star}(V_\star,V_\star)$
(with product $\bullet_{A_\star}$) is isomorphic
to the $\star$-matrix product algebra $A_\star^{n\times n}$.

\paragraph{Tensor product:}  
Given $ V_\star = A_\star^n$, $W_\star = A_\star^m$, $X_\star = A_\star^l$
and $Y_\star = A_\star^p$, one can show by similar calculations 
that the tensor product $L^\prime\obultimes_{A_\star} L\in\hom_{A_\star}(V_\star\otimes_{A_\star} W_\star, X_\star\otimes_{A_\star} Y_\star)$ of any $L\in \hom_{A_\star}(V_\star,X_\star)$ and $L^\prime \in \hom_{A_\star}(W_\star,Y_\star)$
is given by the components
\begin{flalign}\label{eqn:obultimes}
\ev_{A_\star}^{}\big((L \obultimes_{A_\star} L^\prime) \otimes_{A_\star} (e_i \otimes_{A_\star} e_j)\big) 
= (e_k \otimes_{A_\star} e_r) \,({L^k}_i \,\star\, {{L^\prime}\,^r}_j) ~.
\end{flalign} 
Hence the isomorphism \eqref{eqn:homAisomatrix} sends the tensor product operation 
$\obultimes_{A_\star}$ to the $\star$-outer product
\begin{flalign}\label{eqn:oblultimesmatrix}
\ostartimes : A_\star^{l \times n}\otimes_{A_\star} A_\star^{p\times m} \longrightarrow A_{\star}^{(l\,p)\times (n\,m)}~,\quad 
({L^k}_i ) \otimes_{A_\star} ({{L^\prime}\,^r}_j)\longmapsto ({L^k}_i \,\star\, {{L^\prime}\,^r}_j) ~.
\end{flalign}


\subsection{\label{subsec:formhoms}Form-valued homomorphism bundles}
As we shall see in more detail in the next sections, many
homomorphisms in differential geometry 
are valued in the exterior algebra of differential forms $\Omega_\star^\sharp$ on $A_\star$,
i.e.\ they are maps  $L\in
\hom_{A_\star}(V_\star,W_\star\otimes_{A_\star}\Omega_\star^\sharp)$
for some modules $V_\star$ and $W_\star$.
The differential forms $\Omega_\star^\sharp$ on $A_\star$ are obtained 
by twisting, with respect to the cochain twist $F\in H\otimes H$,
the differential forms $\Omega^\sharp(M)$ on the underlying classical manifold $M$:
As vector spaces $\Omega_\star^\sharp = \Omega^\sharp(M)$, while the product on $\Omega_\star^\sharp$ is given by the $\star$-exterior product
\begin{flalign}
\wedge_\star :=\wedge \circ F^{-1} : \Omega_\star^p\otimes_\star \Omega_\star^q \longrightarrow \Omega_\star^{p+q}~.
\end{flalign}
The relevant $H$-action on $\Omega^\sharp(M)$ is given by the Lie derivative of vector fields on forms.
Similarly to \eqref{eqn:flip}, 
the (graded) noncommutativity of the $\star$-exterior product is 
controlled by the $R$-matrix,
\begin{flalign}
\omega \,\wedge_\star\, \omega^\prime = (-1)^{\vert \omega\vert \, \vert \omega^\prime\vert} \, \big(R_F^{(2)} \ra \omega^\prime\big)\,\wedge_\star\,\big(R_F^{(1)} \ra \omega\big)~,
\end{flalign}
for all homogeneous forms $\omega,\omega^\prime \in \Omega_\star^\sharp$.
Nonassociativity is controlled as in \eqref{eqn:naA} by the associator
\begin{flalign}
\big(\omega \,\wedge_\star\, \omega^\prime\big) \,\wedge_\star\, \omega^{\prime\prime} = 
(\phi_F^{(1)} \ra \omega) \,\wedge_\star\, \big((\phi_F^{(2)} \ra \omega^\prime) \,\wedge_\star\, (\phi_F^{(3)} \ra \omega^{\prime\prime})\big)~,
\end{flalign}
for all $ \omega, \omega^\prime, \omega^{\prime\prime} \in \Omega_\star^\sharp $.
The differential 
\begin{flalign}
\dd : \Omega_\star^{p}\longrightarrow \Omega^{p+1}_\star
\end{flalign}
on $\Omega^\sharp_\star$ is given by the ordinary de Rham exterior derivative
and it satisfies the graded Leibniz rule
\begin{flalign}
\dd(\omega\wedge_\star \omega^\prime) = \dd \omega\wedge_\star
\omega^\prime + (-1)^{\vert \omega\vert}\,\omega\wedge_\star \dd
\omega^\prime ~,
\end{flalign}
for all homogeneous forms $\omega,\omega^\prime \in \Omega_\star^\sharp$.
\sk

Because $\Omega_\star^\sharp$ is a graded $H_F$-module algebra 
and not only an $H_F$-module $A_\star$-bimodule, the 
modules of homomorphisms 
$\hom_{A_\star}(V_\star,W_\star\otimes_{A_\star}\Omega_\star^\sharp)$ 
may be equipped with additional structures, which we shall now briefly describe.
\sk
For this, we introduce the notation
\begin{flalign}
V_\star^\sharp := V_\star\otimes_{A_\star} \Omega_\star^\sharp
\end{flalign}
to denote the tensor product of the module $V_\star$ with the module of differential forms $\Omega_\star^\sharp$.
A generic element in $V_\star^\sharp$ is of the form $e_i\otimes_{A_\star}\omega^i$,
where $\omega^i \in\Omega_\star^\sharp$.
Notice that $V_\star^\sharp$ is a graded module, with $V_\star^p = V_\star\otimes_{A_\star} \Omega_\star^p$.
Because $\Omega_\star^\sharp$ is a graded $H_F$-module algebra,
$V_\star^\sharp$ is moreover a graded $H_F$-module $\Omega_\star^\sharp$-bimodule
with left and right $\Omega_\star^\sharp$-action given by the
$\star$-exterior product, i.e.
\begin{subequations}
\begin{flalign}
(e_i\otimes_{A_\star}\omega^i)\wedge_\star \omega^\prime &:= 
e_i \otimes_{A_\star} \big( \omega^i \wedge_\star  \omega^\prime \big)~,\\[4pt]
\omega^\prime\wedge_\star (e_i\otimes_{A_\star}\omega^i) &:= 
e_i \otimes_{A_\star}\big(\omega^\prime \wedge_\star \omega^i\big)~,
\end{flalign}
\end{subequations}
for all  $\omega^i ,\omega^\prime\in \Omega_\star^\sharp$. (Notice that
this definition uses $H_F$-invariance of the basis $e_i$.)
\sk

We shall now show that the module of homomorphisms
$\hom_{A_\star}(V_\star,W_\star\otimes_{A_\star}\Omega_\star^\sharp)$ 
is isomorphic (as an $H_F$-module $A_\star$-bimodule) to the module
$\hom_{\Omega_\star^\sharp} (V_\star^\sharp,W_\star^\sharp)$
of weak right $\Omega^\sharp_\star$-linear maps, which is characterized by the condition (compare with \eqref{eqn:weakright})
\begin{flalign}\label{eqn:omegalin}
\ev_{F}^{}\Big(L \otimes_{\star} \big((e_i \otimes_{A_\star} \omega^i)
\,\wedge_\star\, \omega^\prime\big)\Big) =
\ev_{F}^{}\Big((\phi_F^{(-1)} \ra L) \otimes_{\star} \big(e_i
\otimes_{A_\star} (\phi_F^{(-2)} \ra\omega^i) \big)\Big)\,\wedge_\star (\phi_F^{(-3)} \ra \omega^\prime)~,
\end{flalign}
for all $\omega^i ,\omega^\prime\in \Omega_\star^\sharp$. In fact, following the same arguments
as before, we use the bases of $V_\star= A_\star^n$ and  $W_\star = A_\star^m$
to show that there is an isomorphism of $H_F$-module $\Omega_\star^\sharp$-bimodules
\begin{flalign}\label{eqn:homOmegaisomatrix}
\hom_{\Omega_\star^\sharp}(V^\sharp_\star,W^\sharp_\star) \longrightarrow  {\Omega_\star^\sharp}^{m\times n}~,\quad 
L\longmapsto ({L^{j}}_{i})~.
\end{flalign}
The matrix coefficients are defined by
\begin{flalign}
\ev_{F}^{} \big(L \otimes_{\star} (e_i \otimes_{A_\star} 1)\big) = e_j\otimes_{A_\star} {L^j}_i ~,
\end{flalign}
where $1\in A_\star\subseteq \Omega_\star^\sharp$ is the unit element. Any element
 $L\in \hom_{A_\star}(V_\star,W_\star\otimes_{A_\star}\Omega_\star^\sharp)$
has exactly the same expansion in the bases of $V_\star$ and $W_\star$, hence we can define
an isomorphism
\begin{flalign}\label{eqn:liftiso}
(\,\cdot\,)^{\sharp} :  \hom_{A_\star}(V_\star,W_\star\otimes_{A_\star}\Omega_\star^\sharp)\longrightarrow
\hom_{\Omega_\star^\sharp}(V^\sharp_\star,W^\sharp_\star) 
\end{flalign}
by going via the matrix representations.
\sk

Given $ V_\star = A_\star^n$, $W_\star = A_\star^m$ and $X_\star = A_\star^l$,  
we use the isomorphisms \eqref{eqn:liftiso} and \eqref{eqn:homOmegaisomatrix} to define a composition operation
\begin{subequations}\label{eqn:formcomp}
\begin{flalign}
\bullet_{A_\star} :\hom_{A_\star}(W_\star,X_\star\otimes_{A_\star} \Omega_\star^\sharp) \otimes_{A_\star}
 \hom_{A_\star}(V_\star,W_\star\otimes_{A_\star} \Omega_\star^\sharp)\longrightarrow
\hom_{A_\star}(V_\star,X_\star\otimes_{A_\star} \Omega_\star^\sharp)
\end{flalign}
in terms of the $\wedge_\star$-matrix product
\begin{flalign}
\wedge_\star : {\Omega_\star^\sharp}^{l\times m} \otimes_{A_\star} {\Omega_\star^\sharp}^{m\times n}
\longrightarrow {\Omega_\star^\sharp}^{l\times n}~,\quad ({{L^\prime}\,^k}_j) \otimes_{A_\star} ({{L}^j}_i)\longmapsto
({{L^\prime}\,^k}_j \,\wedge_\star\, {L^j}_i)~.
\end{flalign}
\end{subequations}
Given $ V_\star = A_\star^n$, $W_\star = A_\star^m$, $X_\star = A_\star^l$
and $Y_\star = A_\star^p$, we define a tensor product operation
\begin{subequations}\label{eqn:formotimes}
\begin{flalign}
\obultimes_{A_\star} : \hom_{A_\star}(V_\star, X_\star\otimes_{A_\star} \Omega_\star^\sharp) \otimes_{A_\star}
\hom_{A_\star}(W_\star, Y_\star\otimes_{A_\star} \Omega_\star^\sharp)\longrightarrow
\hom_{A_\star}\big(V_\star\otimes_{A_\star} W_\star, (X_\star
\otimes_{A_\star} Y_\star)\otimes_{A_\star} \Omega_\star^\sharp \big)
\end{flalign}
in terms of the $\wedge_\star$-outer product
\begin{flalign}
\ostartimes : {\Omega_\star^\sharp}^{l \times n}\otimes_{A_\star} {\Omega_\star^\sharp}^{p\times m} 
\longrightarrow {\Omega_\star^\sharp}^{(l\,p)\times (n\,m)}~,\quad 
({L^k}_i ) \otimes_{A_\star} ({{L^\prime}\,^r}_j)\longmapsto ({L^k}_i \,\wedge_\star\, {{L^\prime}\,^r}_j) ~.
\end{flalign}
\end{subequations}
These operations generalize \eqref{eqn:compmatrix} and \eqref{eqn:oblultimesmatrix}
to form-valued homomorphisms.


\section{Nonassociative connections and curvature\label{sec:NACC}}


\subsection{Connections}
A nonassociative connection on a module $V_\star$
is a linear map $\nabla \in \hom_F^{}(V_\star, V_\star\otimes_{A_\star}\Omega_\star^1)$
which satisfies the Leibniz rule
\begin{flalign}\label{eqn:Leibniz}
\ev_F^{}\big(\nabla\otimes_{\star} (v\,\star\,a)\big) =  
\ev_F^{}\big((\phi_F^{(-1)}\ra \nabla) \otimes_{\star} (\phi_F^{(-2)}\ra v) \big)\,\star\,(\phi_F^{(-3)}\ra a) 
\,+\, v \otimes_{A_\star} \dd a~,
\end{flalign}
for all $v \in V_\star$ and $a \in A_\star$, where $ \dd $ is the exterior derivative 
of the differential calculus $\Omega_\star^\sharp$. 
We denote the space of connections on $V_\star$ by $\mathrm{con}_F^{}(V_\star)$ 
and note that it is an affine space over the 
module of homomorphisms $ \hom_{A_\star}(V_\star,  V_\star\otimes_{A_\star}\Omega_\star^1) $.
\sk

As $V_\star = A_\star^n$ is by assumption a free $A_\star$-bimodule,
we can describe any connection $\nabla\in \mathrm{con}_F^{}(V_\star)$ 
in terms of its coefficients ${\Gamma^j}_i\in \Omega^1_\star$ 
defined by
\begin{flalign}\label{eqn:con}
\ev_F^{}(\nabla \otimes_\star e_i) =: e_j \otimes_{A_\star} {\Gamma^j}_i~.
\end{flalign}
Using \eqref{eqn:Leibniz}, after a short calculation we obtain
\begin{flalign}\label{eqn:conv}
\ev_F^{}(\nabla \otimes_\star v) 
= e_i \otimes_{A_\star} \big(\dd v^i \,+\, {\Gamma^i}_j \,\star\, v^j\big) ~,
\end{flalign}
for all $v = e_i\,v^i\in V_\star$.
\sk

As $\mathrm{con}_F^{}(V_\star)\subseteq \hom_F^{}(V_\star,V_\star\otimes_{A_\star}\Omega^1_\star)$
is an affine subspace, we can act with any $h\in H_F$ on a connection $\nabla$
and obtain an element $h\ra \nabla \in \hom_F^{}(V_\star,V_\star\otimes_{A_\star}\Omega^1_\star)$,
which however in general does not lie in $\mathrm{con}_F^{}(V_\star)$: In contrast to the Leibniz rule
\eqref{eqn:conv}, $h\ra \nabla $ satisfies
\begin{flalign}
\ev_F^{}\big((h\ra \nabla) \otimes_\star v\big) 
= e_i \otimes_{A_\star} \big( \epsilon_F^{}(h)\, \dd v^i \,+\, (h\ra {\Gamma^i}_j ) \,\star\, v^j\big) ~,
\end{flalign}
for all $v = e_i\,v^i\in V_\star$. In particular, if $h\in H_F$ satisfies $\epsilon_F(h) =1$ then
$h\ra \nabla\in \mathrm{con}_F^{}(V_\star)$, while if $\epsilon_F(h)=0$ then
$h\ra \nabla\in \hom_{A_\star}(V_\star,V_\star\otimes_{A_\star}\Omega_\star^1)$.
\sk

Similarly to the case of homomorphisms \eqref{eqn:liftiso}, we can lift
connections $\nabla \in \mathrm{con}_F^{}(V_\star)$ to linear maps
$\nabla^\sharp \in \mathrm{end}_F^{}(V_\star^\sharp)$, which then 
satisfy the condition
\begin{flalign}\label{eqn:conomegae}
\ev_F^{}\big(\nabla^\sharp \otimes_\star (e_i \otimes_{A_\star} \omega^i)\big) 
= e_i \otimes_{A_\star} \big( \dd \omega^i \,+\, {\Gamma^i}_j \,\wedge_\star\, \omega^j \big)~,
\end{flalign}
for all $\omega^i \in \Omega_\star^\sharp$. Notice that \eqref{eqn:conomegae}
implies the graded Leibniz rule
\begin{flalign}\label{eqn:cononforms}
\ev_F^{}\big(\nabla^\sharp \otimes_\star (s \,\wedge_\star\, \omega^\prime) \big) = 
\ev_F^{}\big((\phi_F^{(-1)} \ra \nabla^\sharp) \otimes_\star (\phi_F^{(-2)} \ra s) \big)\,\wedge_\star\,
 (\phi_F^{(-3)} \ra \omega^\prime) \,+\, (-1)^{\vert s \vert} \, s\,\wedge_\star\, \dd  \omega^\prime~,
\end{flalign}
for all homogeneous forms
$s = e_i \otimes_{A_\star}\omega^i \in V_\star^\sharp$ and $\omega^\prime\in \Omega_\star^\sharp$.


\subsection{Connections on tensor products} 
Given $V_\star = A_\star^n $ and $W_\star = A_\star^m$,
together with connections $\nabla\in \mathrm{con}_F^{}(V_\star)$
and $\nabla^\prime\in\mathrm{con}_F^{}(W_\star)$,
we can construct a connection on $V_\star\otimes_{A_\star}W_\star$
by taking their sum $\nabla \obulplus_F^{} \nabla^\prime$, see~\cite[Section~4.2]{Barnes:2015}
for details.
In terms of the coefficients ${\Gamma^k}_i ,{{\Gamma^\prime}\,^l}_j\in \Omega_\star^1$,
the sum of connections takes a simple form and it is specified
by the coefficients
\begin{flalign}\label{eqn:sumcomponents}
\ev_F^{}\big((\nabla \obulplus_F^{} \nabla^\prime) \otimes_\star (e_i \otimes_{A_\star} e_j)\big) = 
(e_k \otimes_{A_\star} e_l) \otimes_{A_\star} \big({\Gamma^k}_i  ~{\delta^l}_j \,+\, {\delta^k}_i~{{\Gamma^\prime}\,^l}_j\big)~.\end{flalign}
On a generic element $v\otimes_{A_\star} w = e_i \otimes_{A_\star} e_j \,(v^i\star w^j)\in V_\star\otimes_{A_\star}W_\star$,
the sum of connections acts as
\begin{multline}
\ev_F^{}\big((\nabla \obulplus_F^{} \nabla^\prime) \otimes_\star (v \otimes_{A_\star} w)\big) =\\
(e_k \otimes_{A_\star}e_l) \otimes_{A_\star} \big(\dd(v^k \,\star\, w^l) \,+\, {\Gamma^k}_i \wedge_\star(v^i\,\star\,w^l) \,+\, {{\Gamma^\prime}\,^l}_j \wedge_\star(v^k\,\star\,w^j) \big)~.
\end{multline}

The sum of connections can be consistently extended to tensor products of finitely many modules by inductively using \eqref{eqn:sumcomponents}.
For example, given $V_\star = A_\star^n $, $W_\star = A_\star^m$ and $X_\star= A_\star^l$,
together with connections $\nabla\in \mathrm{con}_F^{}(V_\star)$, $\nabla^\prime\in\mathrm{con}_F^{}(W_\star)$
and $\nabla^{\prime\prime}\in \mathrm{con}_F^{}(X_\star)$,
then $(\nabla\obulplus_F^{} \nabla^\prime)\obulplus_F^{} \nabla^{\prime\prime}\in 
\mathrm{con}_F^{}((V\otimes_{A_\star} W_\star)\otimes_{A_\star}X_\star)$
is specified by the connection coefficients
\begin{multline}
\ev_F^{}\Big(\big((\nabla \obulplus_F^{} \nabla^\prime)\obulplus_F^{}\nabla^{\prime\prime}\big) \otimes_\star \big((e_i \otimes_{A_\star} e_j ) \otimes_{A_\star} e_k\big)\Big) =\\
\big((e_{i^\prime} \otimes_{A_\star} e_{j^\prime})\otimes_{A_\star} e_{k^\prime} \big) 
\otimes_{A_\star} \big({\Gamma^{i^\prime}}_i ~{\delta^{j^\prime}}_j~{\delta^{k^\prime}}_k + {\delta^{i^\prime}}_i~
{{\Gamma^\prime}\,^{j^\prime}}_j ~{\delta^{k^\prime}}_k + {\delta^{i^\prime}}_i~{\delta^{j^\prime}}_j~
{{\Gamma^{\prime\prime}}\,^{k^\prime}}_k\big)~.
\end{multline}
Moreover, $(\nabla\obulplus_F^{} \nabla^\prime)\obulplus_F^{} \nabla^{\prime\prime}$ and 
$\nabla \obulplus_F^{} (\nabla^\prime\obulplus_F^{} \nabla^{\prime\prime})$ are related by adjoining the associator
\begin{flalign}
(\nabla\obulplus_F^{} \nabla^\prime)\obulplus_F^{} \nabla^{\prime\prime} = \phi_F^{-1}\circ \big(\nabla \obulplus_F^{} (\nabla^\prime\obulplus_F^{} \nabla^{\prime\prime}) \big)\circ \phi_F~.
\end{flalign}


\subsection{Connections on homomorphism bundles}
Given $V_\star = A_\star^n $ and $W_\star = A_\star^m$,
together with connections $\nabla\in \mathrm{con}_F^{}(V_\star)$
and $\nabla^\prime\in\mathrm{con}_F^{}(W_\star)$,
we can construct a connection on $\hom_{A_\star}(V_\star,W_\star)$
by taking their adjoint $ {\DDD}_F(\nabla^\prime, \nabla)$, see~\cite[Section~4.3]{Barnes:2015} for details.
In terms of the coefficients ${\Gamma^k}_i,{{\Gamma^\prime}\,^l}_j\in \Omega_\star^1$,
the adjoint connection takes a simple form:
Denoting by $\{{e_{j}}^{i}\}$ the basis of $\hom_{A_\star}(V_\star,W_\star)$
given by the isomorphism \eqref{eqn:homAisomatrix} and the standard basis
of $A_\star^{m\times n}$, the coefficients of ${\DDD}_F(\nabla^\prime, \nabla)$
are given by
\begin{flalign}\label{eqn:adcon}
\ev_F^{}\big({\DDD}_F(\nabla^\prime, \nabla) \otimes_\star {e_{j}}^i\big) = 
{e_{j^\prime}}^{i^\prime}\otimes_{A_\star} \big({{\Gamma^\prime}\,^{j^\prime}}_{j}~{\delta^{i}}_{i^\prime} -
{\delta^{j^\prime}}_{j} ~ {\Gamma^{i}}_{i^\prime}\big)~.
\end{flalign}
On a generic element $L = {e_{j}}^{i}\,{L^j}_{i}\in \hom_{A_\star}(V_\star,W_\star)$,
the adjoint connection acts as
\begin{flalign}\label{eqn:adcomponents}
\ev_F^{}\big({\DDD}_F(\nabla^\prime, \nabla) \otimes_\star L\big)
={e_{j^\prime}}^{i^\prime}\otimes_{A_\star} \big(\dd {L^{j^\prime}}_{i^\prime} + 
{{\Gamma^\prime}\,^{j^\prime}}_j \star {L^{j}}_{i^\prime} 
- (R_F^{(2)}\ra {L^{j^\prime}}_{i}) \star (R^{(1)}_F\ra {\Gamma^{i}}_{i^\prime})\big)~,
\end{flalign}
where in the last term we have used the $R$-matrix to rearrange
the term ${\Gamma^{i}}_{i^\prime} \star {L^{j^\prime}}_{i}$ so
that $\star$-matrix multiplication is obvious.
\sk

The adjoint connection extends to
form-valued homomorphisms 
$L\in \hom_{A_\star}(V_\star,W_\star\otimes_{A_\star}\Omega_\star^\sharp)$.
The resulting expression
\begin{flalign}\label{eqn:adcomponentsform}
\ev_F^{}\big({\DDD}_F(\nabla^\prime, \nabla) \otimes_\star L\big)
={e_{j^\prime}}^{i^\prime}\otimes_{A_\star} \big(\dd {L^{j^\prime}}_{i^\prime} + 
{{\Gamma^\prime}\,^{j^\prime}}_j \, \wedge_\star\,  {L^{j}}_{i^\prime} 
- (-1)^{\vert L\vert} \, (R_F^{(2)}\ra {L^{j^\prime}}_{i}) \,\wedge_\star\, (R^{(1)}_F\ra {\Gamma^{i}}_{i^\prime})\big)
\end{flalign}
is very similar to \eqref{eqn:adcomponents}
whereby we simply replace $\star$-products by $\wedge_\star$-products
and include a degree-dependent sign factor in front of the last term.
\sk

As an important example, let us consider the dual module
$V_\star^\vee :=   \hom_{A_\star}(V_\star,A_\star)$ of $V_\star = A_\star^n$.
Following the notations used above, we denote the basis
of the dual module by $\{e^i\}$, i.e.\ with an upper index.
A generic element in $V_\star^\vee$ is thus of the form
$L = e^i\,L_i$ with $L_i\in A_\star$.
Given now any connection $\nabla\in\mathrm{con}_F^{}(V_\star)$, 
we can use the differential $\dd : A_\star\to\Omega_\star^1$ as
a connection on $A_\star$, and define
a connection on $V_\star^\vee$ by taking the adjoint connection 
$\nabla^{\vee} := {\DDD}_F(\dd,\nabla)$. Because $\dd$ does not have any nontrivial 
connection coefficients, the general expression \eqref{eqn:adcomponents}
implies that the dual connection acts on $L = e^i\,L_i\in V_\star^\vee$ as
\begin{flalign}\label{eqn:forconjugatefield}
\ev_F^{}\big(\nabla^\vee \otimes_\star L\big) = e^{i^\prime} 
\otimes_{A_\star} \big(\dd\, L_{i^\prime} \,-\, (R_F^{(2)} \ra L_{i}) \,\star\, (R_F^{(1)} \ra {{\Gamma}^{i}}_{i^\prime})\big)~.
\end{flalign}


\subsection{Curvature}
The curvature of a connection $\nabla\in\mathrm{con}_F^{}(V_\star)$
is given by the graded $R$-matrix commutator 
\begin{flalign}\label{eqn:curv}
R(\nabla):= \mbox{$\frac{1}{2}$} \,[\nabla^\sharp, \nabla^\sharp]_{F}^{} := 
\mbox{$\frac{1}{2}$}\, \big(\nabla^\sharp \bullet_F \nabla^\sharp + (R_F^{(2)} \ra \nabla^\sharp) \bullet_F (R_F^{(1)} \ra \nabla^\sharp)\big)~
\end{flalign}
of its lift $\nabla^\sharp\in\mathrm{end}_F^{}(V_\star^\sharp)$ defined in \eqref{eqn:conomegae}.
Due to the graded Leibniz rule \eqref{eqn:cononforms}, it follows that $R(\nabla) 
\in \hom_{A_\star}(V_\star,V_\star\otimes_{A_\star} \Omega_\star^2)$ is a homomorphism valued in 2-forms.
The coefficients of the curvature are given by
\begin{subequations}\label{eqn:curve}
\begin{flalign}
\ev_{A_\star}^{}\big(R(\nabla) \otimes_{A_\star} e_i  \big)
= e_j \otimes_{A_\star} {R^j}_i = e_j \otimes_{A_\star} \big(\dd {\Gamma^j}_i \,+\, \mbox{$\frac{1}{2}$}\, {{[\Gamma \,,\, \Gamma]_\star}^j}_i~\big)~,
\end{flalign}
where
\begin{flalign}
{{[\Gamma \,,\, \Gamma]_\star}^j}_{i} := {\Gamma^j}_k \wedge_\star {\Gamma^k}_i + \big(R_F^{(2)} \ra {\Gamma^j}_k\big) \wedge_\star \big(R_F^{(1)} \ra{\Gamma^k}_i \big)~.
\end{flalign}
\end{subequations}
On the sum of connections $\nabla\in \mathrm{con}_F^{}(V_\star)$
and $\nabla^\prime\in\mathrm{con}_F^{}(W_\star)$, the curvature $R(\nabla \obulplus_F^{} \nabla^\prime)$ has the desired additive behavior
\begin{flalign}
\ev_{A_\star}^{}\big(R(\nabla \obulplus_F^{} \nabla^\prime) \otimes_{A_\star} (e_i \otimes_{A_\star} e_j)\big) = 
(e_k \otimes_{A_\star} e_l) \otimes_{A_\star} \big({R^k}_i  ~{\delta^l}_j \,+\, {\delta^k}_i~{{R^\prime}\,^l}_j\big)~.
\end{flalign}

The Bianchi tensor of a connection $\nabla\in\mathrm{con}_F^{}(V_\star)$ is defined by
acting with the adjoint connection on the curvature using \eqref{eqn:adcomponentsform} to get
\begin{flalign}\label{eqn:bianchi}
\mathrm{Bianchi}(\nabla) &:= \ev_F^{}\big({\DDD}_F(\nabla,\nabla) \otimes_{\star}R(\nabla)\big) ~.
\end{flalign}
By definition, it follows that
$\mathrm{Bianchi}(\nabla) \in \hom_{A_\star}(V_\star, V_\star\otimes_{A_\star}\Omega_\star^3)$
is a homomorphism valued in 3-forms. Using \eqref{eqn:adcomponentsform}
we find
\begin{subequations}
\begin{flalign}
\ev_{A_\star}^{}\big(\mathrm{Bianchi}(\nabla) \otimes_{A_\star} e_i  \big)
= e_j \otimes_{A_\star} {{\mathrm{Bianchi}}^j}_i = e_j \otimes_{A_\star} \big(\dd {R^j}_i + {{[\Gamma, R]_\star}^j}_i\big)~,
\end{flalign}
where
\begin{flalign}
{{[\Gamma, R]_\star}^j}_i := {{\Gamma}^j}_k \,\wedge_\star\,{R^k}_i\,-\, \big(R_F^{(2)} \ra {R^j}_k\big) \,\wedge_\star\, \big(R_F^{(1)} \ra {{\Gamma}^k}_i\big)~.
\end{flalign}
\end{subequations}
An interesting consequence of the noncommutativity and 
nonassociativity of $ A_\star $ (which is controlled by the $R$-matrix and associator)
is that in general the Bianchi tensor does not vanish, i.e. the Bianchi identity is generally violated. However, for trivial $R$-matrix and associator
we recover the usual Bianchi identity in classical differential geometry 
for any connection $\nabla$.


\section{Nonassociative field theory\label{sec:NAFT}}


\subsection{Yang-Mills theory}
Let $M$ be an oriented $m$-dimensional
manifold equipped with an $H$-invariant
Riemannian or Lorentzian metric. Then
the classical Hodge operator $\ast_M^{} : \Omega^p(M)\to \Omega^{m-p}(M)$
is $H$-equivariant, i.e.\ $\ast_{M}^{}\circ (h\ra \, \cdot\, ) = (h\ra \,\cdot\,) \circ \ast_{M}^{}$ for all
$h\in H$. We equip the deformed differential forms with the same Hodge operator,
leading to an $H_F$-equivariant map
\begin{flalign}
\ast_{M}^{} : \Omega_\star^p \longrightarrow \Omega_\star^{m-p}~.
\end{flalign}

Given any module $V_\star = A_\star^n$ and any connection
$\nabla \in \mathrm{con}_F^{}(V_\star)$, 
let $\mathcal{L}(\nabla) \in
\hom_{A_\star}^{}(V_\star,V_\star\otimes_{A_\star}\Omega_\star^m)$
be the homomorphism valued in top-forms which is
given by the components
\begin{flalign}
{\mathcal{L}^j}_i = \mbox{$\frac{1}{2}$}~{F^j}_k \,\wedge_\star\, \ast_M^{}  {F^k}_i~,
\end{flalign}
where as usual we denote the curvature of a gauge connection
by ${F^j}_i = \dd {\Gamma^j}_i \,+\, \frac{1}{2}\, {{[\Gamma \,,\, \Gamma]_\star}^j}_i$.
The action functional for Yang-Mills gauge theory is given by tracing and integrating $\mathcal{L}(\nabla) $, i.e.\ 
\begin{flalign}\label{eqn:YMaction}
S_{\mathrm{YM}}(\nabla) := \int_M \,
\mathrm{Tr}\big(\mathcal{L}(\nabla) \big) = \frac{1}{2}\, \int_M \,
{F^j}_k \,\wedge_\star\, \ast_M^{}  {F^k}_j ~.
\end{flalign}
We shall now show that, under certain natural conditions on the twist $F\in H\otimes H$
and the connection $\nabla$, the Yang-Mills action \eqref{eqn:YMaction} is real-valued. 
\sk

The first condition is that $F$ is Hermitean, i.e.\ it defines a Hermitean star-product
on $A_\star$. This means that $(a\star b)^\ast = b^\ast\star a^\ast$, where
${}^\ast$ denotes the involution given by 
pointwise complex conjugation of functions on $M$. This is clearly the case for
Examples \ref{ex:abelian} and \ref{ex:nonassociative}.
We extend the involution ${}^\ast$ on $A_\star$ to
a graded involution on the differential forms $\Omega_\star^\sharp$ by setting
\begin{flalign}\label{eqn:omegainv}
(\omega \wedge_\star \omega^\prime)^\ast = (-1)^{\vert \omega \vert\,  \vert \omega^\prime \vert}~ 
{\omega^\prime}\,^\ast \wedge_\star \omega^\ast ~,\quad
(\dd\omega)^\ast = \dd\omega^\ast ~,
\end{flalign}
for all homogeneous forms $\omega,\omega^\prime\in \Omega_\star^\sharp$.
\sk

The second condition is that $\nabla$ is unitary, i.e.\
the corresponding connection coefficients satisfy 
\begin{flalign}
{{\Gamma^j}_i}^\ast = - {\Gamma^{i}}_j~.
\end{flalign}
Using \eqref{eqn:curve} one easily shows that the curvature of
a unitary connection is an anti-Hermitean matrix, i.e.\ 
\begin{flalign}
{{F^j}_i}^\ast = - {F^{i}}_j~.
\end{flalign}

The third condition is the graded $2$-cyclicity property
\begin{flalign}
\int_M \, \omega \,\wedge_\star\, \omega^\prime = (-1)^{\vert \omega
  \vert\,  \vert \omega^\prime \vert}\, \int_M \, \omega^\prime \,\wedge_\star\, \omega~,
\end{flalign}
for all homogeneous forms $ \omega, \omega^\prime \in
\Omega_\star^\sharp $. This property holds for Abelian twists,
as in Example \ref{ex:abelian}, and also for the nonassociative
deformation of Example \ref{ex:nonassociative}, see
\cite{Mylonas:2013jha}.
\sk

The first two conditions imply that the complex conjugate of the action \eqref{eqn:YMaction}
can be simplified as
\begin{flalign}\label{eqn:tmpYMreal}
{S_{\mathrm{YM}}(\nabla)}^\ast =  \frac{1}{2}\, \int_M \, \big({{F^j}_k \,\wedge_\star\, \ast_{M}^{}  {F^k}_j}\big)^\ast
= \frac{1}{2}\, \int_M \, \ast_{M}^{} { {F^k}_j}^\ast \,\wedge_\star\, {{F^j}_k}^\ast  =
\frac{1}{2}\, \int_M \, \ast_{M}^{} {F^j}_k \,\wedge_\star\, {F^k}_j ~,
\end{flalign}
where in the second step we have also used compatibility between the
Hodge operator and the complex conjugation involution.
The third condition then implies that we can interchange the two terms in the last equality of \eqref{eqn:tmpYMreal},
and hence find that the noncommutative and nonassociative Yang-Mills action
is real, i.e.
\begin{flalign}
{S_{\mathrm{YM}}(\nabla)}^\ast = S_{\mathrm{YM}}(\nabla)~.
\end{flalign}
In particular, the noncommutative and nonassociative Yang-Mills action
\eqref{eqn:YMaction} is real-valued for all unitary connections
in Examples \ref{ex:abelian}  and  \ref{ex:nonassociative}.


\subsection{Einstein-Cartan gravity}
The field content of Einstein-Cartan gravity is 
a spin connection $\nabla$ and a vielbein field $E$.
Let $M$ be an oriented $m$-dimensional
manifold which admits a trivial Dirac spinor bundle
\begin{flalign}
S = M\times \bbC^{2^{\lfloor\frac{m}{2}\rfloor}} \longrightarrow M~.
\end{flalign}
We denote the module of sections of the 
spinor bundle by $V :={\mit\Gamma}^\infty(S) =  A^{2^{\lfloor\frac{m}{2}\rfloor}}$.
\sk

Without loss of
generality, here we can take $H = U\mathrm{Vec}(M)$
to be the Hopf algebra of all infinitesimal diffeomorphisms of
$M$. Then given any cochain twist $F \in H\otimes H$, we twist
$A=C^\infty(M)$ to a noncommutative and nonassociative algebra $A_\star$
and $V$ to an $H_F$-module $A_\star$-bimodule $V_\star = A_\star^{2^{\lfloor\frac{m}{2}\rfloor}}$.
\sk

A spin connection on $V_\star$ is a connection 
$\nabla\in\mathrm{con}_F^{}(V_\star)$ for which
the coefficients take the special form
\begin{flalign}\label{eqn:spinconnection}
{\Gamma^j}_i =  \mbox{$\frac{1}{4}$}\, \omega^{ab}\,{{\gamma_{ab}}^j}_i~,
\end{flalign}
where $\omega^{ab} \in \Omega_\star^1$ is antisymmetric in $ab$
and $\gamma_{ab} = \frac{1}{2}\,[\gamma_{a} ,\gamma_{b}]$
is given by the commutator of the gamma-matrices $\gamma_a$; here
the indices $a,b,\dots$ run from $1$ to $m$, the dimension of $M$,
while $i,j,\dots$ run from $1$ to $2^{\lfloor\frac{m}{2}\rfloor}$, 
the rank of the Dirac spinor bundle $S$.
The curvature \eqref{eqn:curve} of a spin connection 
can be computed with some standard gamma-matrix algebra and it reads as
\begin{flalign}
{R^j}_i  =  \mbox{$\frac{1}{4}$}\, R^{ab} \,{{\gamma_{ab}}^j}_i = 
\mbox{$\frac{1}{4}$} \,\big( \dd \omega^{ab} \,+\,{\omega^a}_c \wedge_\star {\omega}^{cb}\big)\,{{\gamma_{ab}}^j}_i~,
\end{flalign}
where the $c$-index was lowered by the flat metric $\eta_{ab}$.
\sk

A vielbein is a homomorphism
$E \in \hom_{A_\star}(V_\star,V_\star\otimes_{A_\star}\Omega_\star^1)$
valued in 1-forms
for which the coefficients take the special form
\begin{flalign}
{E^j}_i =  E^a \,{{\gamma_a}^j}_i ~,
\end{flalign}
where $E^a\in \Omega_\star^1$. 
\sk

Let us assume for the moment that the dimension $m$ of $M$ is even.
We propose the noncommutative and nonassociative generalization
of the Einstein-Cartan action functional given by
\begin{flalign}\label{eqn:ECactioneven}
S_{\mathrm{EC}}^{\mathrm{even}}(\nabla,E) :=\int_M \, \Big( E_{\mathrm{left}}^{a_1\cdots a_{\frac{m}{2}-1}} \wedge_\star R^{a_{\frac{m}{2}}a_{\frac{m}{2}+1}}\Big)\wedge_\star E_{\mathrm{right}}^{a_{\frac{m}{2}+2}\cdots a_{m}}~\epsilon_{a_1\cdots a_m}~~,
\end{flalign}
where $\epsilon_{a_1\cdots a_m}$ is the antisymmetric tensor and 
\begin{subequations}\label{eqn:Elr}
\begin{flalign}
E_{\mathrm{left}}^{a_1\cdots a_{k}} &:= \big(\cdots \big((E^{[a_1}
\wedge_\star E^{a_2})\wedge_\star E^{a_3}\big)\cdots \big)
\wedge_\star E^{a_{k}]} ~,\\[4pt]
E_{\mathrm{right}}^{a_1\cdots a_{k}} &:= E^{[a_1}\wedge_\star \big(  
\cdots \big(E^{a_{k-2}}\wedge_\star (E^{a_{k-1}}\wedge_\star
E^{a_k]})\big)\cdots \big) ~,
\end{flalign}
\end{subequations}
is the $\wedge_\star$-product of $k$ vielbeins in $\Omega^k_\star$ with special bracketing conventions
and totally antisymmetrized (with weight $1$) in the indices $a_1\cdots a_k$.
This choice of bracketing allows us to show that the Einstein-Cartan action \eqref{eqn:ECactioneven}
is real-valued, under similar assumptions as for the Yang-Mills action.
\sk

Let us now assume that the twist $F$ is Hermitean
and further demand the reality conditions
\begin{flalign}
{\omega^{ab}}^\ast =-\omega^{ba} = \omega^{ab}~,\quad {E^a}^\ast = E^a~,
\end{flalign}
for the spin connection and vielbein.
As a consequence, we obtain 
\begin{flalign}
{R^{ab}}^\ast =- R^{ba} = R^{ab}~,\quad {E_{\mathrm{left}}^{a_1\cdots a_{k}}}^\ast =  E_{\mathrm{right}}^{a_1\cdots a_k}~.
\end{flalign}
The complex conjugate of the action \eqref{eqn:ECactioneven} can now be simplified as
\begin{flalign}
\nn {S_{\mathrm{EC}}^{\mathrm{even}}(\nabla,E)}^\ast
&= (-1)^{\frac{m}{2}-1} \, \int_M \, E_{\mathrm{left}}^{a_{\frac{m}{2}+2}\cdots a_{m}} \wedge_\star \Big( R^{a_{\frac{m}{2}}a_{\frac{m}{2}+1}} \wedge_\star E_{\mathrm{right}}^{a_1\cdots a_{\frac{m}{2}-1}}\Big)~\epsilon_{a_1\cdots a_m}\\[4pt]
&= \int_M\, E_{\mathrm{left}}^{a_{1}\cdots a_{\frac{m}{2}-1}} \wedge_\star \Big( R^{a_{\frac{m}{2}}a_{\frac{m}{2}+1}} \wedge_\star E_{\mathrm{right}}^{a_{\frac{m}{2}+2}\cdots a_{m}}\Big)~\epsilon_{a_1\cdots a_m}~,\label{eqn:tmpECreal}
\end{flalign}
where the sign factor in the first equality 
is due to \eqref{eqn:omegainv}. In the second equality we have reordered 
the indices of $\epsilon_{a_1\cdots a_m}$ by using its total antisymmetry property.
\sk

We further assume the $3$-cyclicity property 
\begin{flalign}
\int_M \, (\omega \wedge_\star \omega^\prime) \wedge_\star \omega^{\prime\prime} = 
\int_M \, \omega \wedge_\star (\omega^\prime \wedge_\star \omega^{\prime\prime} )~,
\end{flalign}
for all
$\omega,\omega^\prime,\omega^{\prime\prime}\in\Omega^\sharp_\star$. This property obviously holds for Abelian twists 
as in Example~\ref{ex:abelian}, because they give strictly associative deformations.
For the nonassociative deformation of Example~\ref{ex:nonassociative} the $3$-cyclicity property is shown in
 \cite{Mylonas:2013jha}.
We can then rebracket the expression after the last equality of \eqref{eqn:tmpECreal}
and find that the noncommutative and nonassociative Einstein-Cartan action in even dimensions
\eqref{eqn:ECactioneven} is real, i.e.
\begin{flalign}
{S_{\mathrm{EC}}^{\mathrm{even}}(\nabla,E) }^\ast = S_{\mathrm{EC}}^{\mathrm{even}}(\nabla,E) ~.
\end{flalign}

In the case of an odd-dimensional manifold $M$, one way to obtain a real-valued
Einstein-Cartan action functional is to modify \eqref{eqn:ECactioneven} as
\begin{multline}\label{eqn:ECactionodd}
 S_{\mathrm{EC}}^{\mathrm{odd}}(\nabla,E) :=\frac{1}{2}\, \int_M \,
 \Big( E_{\mathrm{left}}^{a_1\cdots a_{\frac{m-1}{2}-1}} \wedge_\star R^{a_{\frac{m-1}{2}}a_{\frac{m-1}{2}+1}}\Big)\wedge_\star E_{\mathrm{right}}^{a_{\frac{m-1}{2}+2}\cdots a_{m}}~\epsilon_{a_1\cdots a_m}\\
 \quad +\, \frac{1}{2}\, \int_M \, \Big( E_{\mathrm{left}}^{a_1\cdots a_{\frac{m-1}{2}}} \wedge_\star R^{a_{\frac{m-1}{2}+1}a_{\frac{m-1}{2}+2}}\Big)\wedge_\star E_{\mathrm{right}}^{a_{\frac{m-1}{2}+3}\cdots a_{m}}~\epsilon_{a_1\cdots a_m}~~,
\end{multline}
where in the first line the form degree of $E_{\mathrm{right}}$ is larger by 1 than the
form degree of $E_{\mathrm{left}}$ and vice versa in the second line. 
Under the same assumptions as in the even-dimensional case, 
one can show that the action \eqref{eqn:ECactionodd} is real-valued, i.e.
\begin{flalign}
{S_{\mathrm{EC}}^{\mathrm{odd}}(\nabla,E) }^\ast = S_{\mathrm{EC}}^{\mathrm{odd}}(\nabla,E) ~.
\end{flalign}
In fact, the second term in \eqref{eqn:ECactionodd} is the conjugate of the first term and vice versa.
\sk

In particular, the noncommutative and nonassociative Einstein-Cartan gravity 
action in even dimensions \eqref{eqn:ECactioneven} and in odd
dimensions \eqref{eqn:ECactionodd} is real-valued in Examples \ref{ex:abelian}  and  \ref{ex:nonassociative}.


\acknowledgments{
We thank P.~Aschieri, M.~Fuchs and V.~Kupriyanov for helpful discussions. This work was supported in part by the Action MP1405 QSPACE 
from the European Cooperation in Science and Technology
(COST). G.E.B.\ is a Commonwealth Scholar, funded by the UK government. 
The work of A.S.\ is supported by a Research Fellowship of the Deutsche 
Forschungsgemeinschaft (DFG, Germany). 
The work of R.J.S.\ is supported in part by the Consolidated Grant ST/L000334/1 
from the UK Science and Technology Facilities Council (STFC). 
}


\end{document}